\documentclass[journal,twocolumn]{IEEEtran}

\usepackage{cite}

\ifCLASSINFOpdf
   \usepackage[pdftex]{graphicx}
   \graphicspath{{./Pic/}{../jpeg/}}
   \DeclareGraphicsExtensions{.pdf,.jpeg,.png}
\else
   \usepackage[dvips]{graphicx}
   \graphicspath{{./Pic/}}
   \DeclareGraphicsExtensions{.eps,.jpg}
\fi
\usepackage[cmex10]{amsmath}
\usepackage{amsfonts}
\usepackage{bm}
\usepackage{amssymb}
\usepackage{algorithmic}

\usepackage{array}

\ifCLASSOPTIONcompsoc
  \usepackage[caption=false,font=normalsize,labelfont=sf,textfont=sf]{subfig}
\else
  \usepackage[caption=false,font=footnotesize]{subfig}
\fi
\usepackage{fixltx2e}
\usepackage{url}

\IEEEpeerreviewmaketitle

\begin{document}
%
\title{DOA Estimation of Coherent Signals on Coprime Arrays Exploiting Fourth-Order Cumulants}
%
%
%

\author{Yang~Hu,~\IEEEmembership{}
        Yimin~Liu,~\IEEEmembership{Member,~IEEE,}
        and~Xiqin~Wang,~\IEEEmembership{}
\thanks{Y. Hu et. al. are with the Department
of Electronic Engineering, Tsinghua University,  Beijing, 100084, China.}
\thanks{The work of Y. Liu was supported by the National Natural Science Foundation of China (Grant No. 61571260). Corresponding e-mail: yiminliu@tsinghua.edu.cn.}
}

%
%


\maketitle

\begin{abstract}
This paper considers the problem of direction-of-arrival (DOA) estimation of coherent signals on passive coprime arrays, where we resort to the fourth-order cumulants of the received signal to explore more information. 
A fourth-order cumulant matrix (FCM) is introduced for the coprime arrays.
The special structure of the FCM is combined with the array configuration to resolve the coherent signals. 
Since each sparse array of the coprime arrays is uniform, a series of overlapping identical subarrays can be extracted. Using this property,
we propose a generalized spatial smoothing scheme applied to the FCM. From the smoothed FCM, the DOAs of both the coherent and independent signals can be successfully estimated on the pseudo-spectrum generated by the fourth-order MUSIC algorithm.
To overcome the problem of occasional false peak appearing on the pseudo-spectrum, we use a supplementary sparse array whose inter-sensor spacing is coprime to that of either existing sparse array. From the combined spectrum aided by the supplementary sensors, the false peaks are removed while the true peaks remain. The effectiveness of the proposed methods is demonstrated by simulation examples.
\end{abstract}


\section{Introduction}
A pair of coprime arrays consist of two uniform sparse arrays, from which a virtual uniform linear array (ULA) can be constructed from the spatial differences between any two sensors\cite{Vaidyanathan2011}\cite{Pal2011}. The spatial autocorrelations at all lags are estimated on the virtual ULA. The increased freedom has been used to identify ${\cal O}(MN)$ sources from only ${\cal O}(M+N)$ physical sensors \cite{Zhang2013}. Due to the simplicity of the array configuration, and the ability to resolve much more signals than the number of sensors, coprime arrays have attracted considerable interests in the DOA estimation applications \cite{Tan2013a} \cite{Pakrooh2016}.
In real scenarios, due to multi-path propagation or smart jammers, signals from different DOAs may become partially correlated, or coherent (fully correlated) in the extreme case \cite{Cozzens1994}. The correlated/coherent signals pose a great challenge to the DOA estimation on coprime arrays. Since the spatial autocorrelations are estimated from the sample mean of the sensor-to-sensor signal multiplications, the presence of coherent signals indicates that the spatial autocorrelations contain cross-terms, which strongly affects the structure of the signal subspace. Incorrect extraction of the signal subspace brings about a failed DOA estimation.

The spatial smoothing preprocessing scheme was developed for a physical ULA to resolve coherent signals \cite{Shan1985}. On coprime arrays, such scheme was employed to construct a correlation matrix for the virtual ULA \cite{Pal2010}. However, the scheme cannot eliminate the cross-terms and hence the coherent signal problem is not solved.
Recently, BouDaher et.al. proposed an algorithm to locate coherent targets using active sensing approach on coprime MIMO \cite{BouDaher2015}. But their method cannot be used for the DOA estimation on passive coprime arrays.

The fourth-order (FO) array processing methods were developed for the DOA estimation of non-Gaussian signals \cite{Porat1991} \cite{Chiang1989}. The main interests in using the FO processing cumulants relies on the increased degrees-of-freedom provided by the virtual coarray, and the higher resolution brought by the larger effective aperture \cite{Dogan1995} \cite{Chevalier1999} \cite{Chevalier2005}.
Currently, the FO methods are used in coprime arrays \cite{Shen2016} or nested arrays \cite{Pal2012a} to increase the virtual aperture. However, as the authors stated, their algorithms cannot handle coherent signals. 

In our work, the scenario where the independent and coherent signals coexist is considered. 
We first formulate an FO cumulant matrix (FCM) with a special form, from which the DOA estimation can be carried out by the fourth-order MUSIC (4-MUSIC) algorithm \cite{Porat1991}.
Unfortunately, the FCM cannot be used for DOA estimation of the coherent signals directly.
The particular form of the FCM is combined with the array configuration to resolve coherent signals. Since each sparse array is uniform, a series of overlapping identical subarrays can be extracted. 
Taking one such subarray from each of the sparse arrays, we can build a pair of coprime subarrays. An FCM is inherently defined on such coprime subarrays, whose size is determined by the subarray sensor numbers. On two similar pairs of coprime subarrays, the FCMs share the same structure.
Analogous to the spatial smoothing scheme applied to the correlation matrix of a ULA, we propose a generalized spatial smoothing scheme applied to the FCM. When the smoothed FCM is adopted by the 4-MUSIC algorithm, both the independent and coherent signals can be successfully estimated.

Ocassionally, the pseudo-spectrum generated from the smoothed FCM encounters a false-peak problem. Some false peaks may appear at the directions where none of the true signals resides, interfering the extraction of the true signals. We analyzed the causation of this phenomenon.
To overcome this challenge, a supplementary sparse array can be added, whose inter-sensor spacing is respectively coprime to each of the existing sparse arrays.
On the combined pseudo-spectrums aided by the new array, the false peaks are removed.

This paper is organized as follows. In Section \ref{sec:Signal}, we briefly review the coprime array configuration, and then formulate the signal model. In Section \ref{sec:FCM}, the FO cumulants as well as the form of the FCM is detailed, and the impact of coherent signals on the FCM is investigated. In Section \ref{sec:SS}, a generalized spatial smoothing scheme on the FCM is proposed to resolve coherent signals. Section \ref{sec:FalsePeak} provides a method to remove the false peaks on the pseudo-spectrum. The effectiveness of the new approach is demonstrated in Section \ref{sec:Simulation}. Section \ref{sec:Conclusion} concludes the paper.

Notations: We use lower-case (upper-case) bold characters to denote vectors (matrices). $E\{ \cdot \}$ represents the statistical expectation. $\left( \cdot \right)^T$ and $(\cdot)^H$ respectively denote the transposition and conjugate transposition of a vector or a matrix. $\left( \cdot \right)^*$ is the element-wise complex conjugate. $\otimes$ denotes the Kronecker product. ${\rm rank} (\cdot)$ denotes the rank of a matrix. $\| \bm{x} \|$ is the 2-norm of the vector $\bm{x}$. We use ${\rm diag}(\bm x)$ to denote a diagonal matrix that uses the elements of $\bm x$ as its diagonal elements.

\section{Signal Model} \label{sec:Signal}

\begin{figure}[!t]
\centering
\includegraphics[width=8.5cm]{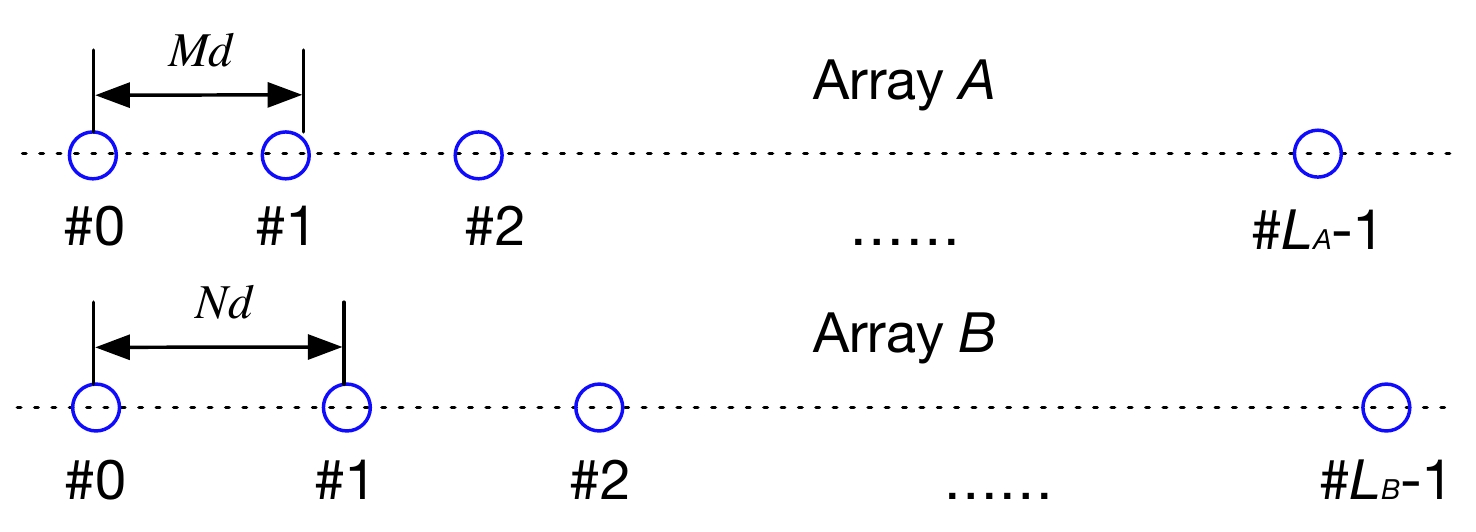} 
\caption{The coprime array configuration} \label{fig:ArrayConfig}
\end{figure}

As illustrated in \figurename \ref{fig:ArrayConfig}, a typical pair of coprime arrays consists of two sparse uniform arrays, denoted by Array $\mathbb A$ and Array $\mathbb B$, respectively. Let $M$ and $N$ be coprime integers, the sensors of the two sparse arrays are located at (with common sensors for both sparse arrays):
\begin{IEEEeqnarray}{rCCl} \label{eq:Manifold} \IEEEyesnumber \IEEEyessubnumber*
 P_A &=& \left\{M i_A d ~ | ~ i_A = 0,1,\ldots,  L_A-1 \right\}&, \\
 P_B &=& \left\{N i_B d ~ | ~ i_B = 0,1,\ldots,  L_B-1 \right\}&.
\end{IEEEeqnarray}
In (\ref{eq:Manifold}), $i_A$ and $i_B$ are the indices of the sensors, the unit inter-sensor spacing $d$ is half-wavelength, $L_A$ and $L_B$ are the number of sensors of Array $\mathbb A$ and $\mathbb B$, respectively. Typically, $L_A \geq N$ and $L_B \geq M$. Denote this pair of coprime arrays by Coarray $\mathbb{AB}$.

Suppose a narrowband signal from the normalized DOA $\theta$ impinges on Coarray $\mathbb{AB}$, the steering vectors for the individual sparse arrays are
\begin{IEEEeqnarray} {rCl}  \IEEEyesnumber \IEEEyessubnumber*
\bm{a} \left(\theta \right) &=& \left[1, e^{jM\theta}, \ldots, e^{jM \left(  L_A-1 \right)\theta} \right]^T, \\
\bm{b} \left(\theta \right) &=& \left[1, e^{jN\theta}, \ldots, e^{jN \left(  L_B-1 \right)\theta} \right]^T.
\end{IEEEeqnarray} 

Let $Q$ narrowband signals impinge on the arrays from the distinct DOAs $\left\{\theta_q\right\}_{q=1}^Q$, and the complex amplitude of the $q$th signal at snapshot time $t$ is $s_q(t)$.
The noise-corrupted measurement vectors on the two sparse arrays are
\begin{IEEEeqnarray} {rCcCl} \label{eq:Receive} \IEEEyesnumber \IEEEyessubnumber*
\bm{y}_{A}\left(t \right) &=& \sum_{q=1}^{Q} \bm{a}(\theta_q) s_q(t) +\bm{w}_A(t) 
&=& \bm{A} \bm{s}\left(t \right)+\bm{w}_A(t), \IEEEeqnarraynumspace \\
\bm{y}_{B}\left(t \right) &=& \sum_{q=1}^{Q} \bm{b}(\theta_q) s_q(t) +\bm{w}_B(t) 
&=& \bm{B} \bm{s}\left(t \right)+\bm{w}_B(t).
\end{IEEEeqnarray}
In (\ref{eq:Receive}), $\bm{w}_A$ and $\bm{w}_B$ are additive noise, $\bm{s}(t)=\left[s_1(t),\ldots,s_Q(t) \right]^T$ is the signal amplitude vector, the matrices $\bm{A}$ and $\bm{B}$ are the collections of steering vectors of Array $\mathbb A$ and $\mathbb B$, respectively
\begin{IEEEeqnarray} {rCl}  \IEEEyesnumber \IEEEyessubnumber*
\bm{A} &=& \left[\bm{a}(\theta_1),\ldots,\bm{a}(\theta_{Q}) \right], \\
\bm{B} &=& \left[\bm{b}(\theta_1),\ldots,\bm{b}(\theta_{Q}) \right].
\end{IEEEeqnarray}

The assumptions on the signals and noises are listed below.
\begin{enumerate}
\item The noise vectors $\bm{w}_A$ and $\bm{w}_B$ in (\ref{eq:Receive}) are zero-mean complex white Gaussian, with arbitrary correlation matrices. The noises are statistically independent of $\bm{s}(t)$.

\item The amplitude of each signal follows a circularly-symmetric complex non-Gaussian distribution. A variety of modulations like QAM or PSK meet this assumption \cite{Porat1991}. Under this assumption, the FO cumulants of the signal are non-zero \cite{Chevalier1999}.

\item Both statistically independent and coherent signals exist. We divide the signals into $G$ groups. The signals in the same group are coherent, and the signals belonging to different groups are statistically independent.

\end{enumerate}

Suppose that there are $Q_g$ coherent signals in the $g$th group ($Q_g=1$ for the independent signal case and $\sum_{g=1}^G Q_g=Q$) with the DOAs $\bm{\theta}_g = [ \theta_{g_1},\ldots,\theta_{g_{Q_g}} ]^T$. Since the amplitudes of coherent signals are linearly dependent \cite{Shan1985}, we can write the group signal amplitude vector by
\begin{IEEEeqnarray} {c} \label{eq:cohcoef}
\bm{s}_g(t) = \bm{\eta}_g \sigma_g(t).
\end{IEEEeqnarray}
In (\ref{eq:cohcoef}), $\bm{\eta}_g=[ \eta_{g_1}, \ldots, \eta_{g_{Q_g}} ]^T$ represents the complex coefficients along the respective propagation paths and hence the elements are non-zero. $\sigma_g(t)$ is a scalar representing the amplitude of the source of the $g$th group at snapshot time $t$.

Because Array $\mathbb A$ and $\mathbb B$ are both sparse and uniform, direction ambiguity exists on the sparse arrays. If a collection of $F$ signals with the DOAs $\{\theta_f\}_{f=1}^F$ satisfies $\theta_f=\theta_1+2\pi m_f /M$ for the distinct non-zero integers $\{ m_f \}_{f=2}^F$, these DOAs are ambiguous on Array $\mathbb A$ because their steering vectors are identical
\begin{IEEEeqnarray} {c}
\bm{a}(\theta_1)=\bm{a}(\theta_2)=\ldots=\bm{a}(\theta_F).
\end{IEEEeqnarray}
If $F$ signals are coherent, and their propagation coefficients are $\{\eta_f\}_{f=1}^F$, it is necessary to assume that
\begin{IEEEeqnarray} {c}
\eta_1+\eta_2+\ldots+\eta_F \neq 0.
\end{IEEEeqnarray}
This assumption guarantees that the ambiguous signals do not vanish on the individual sparse arrays. On the opposite, the collection of coherent signals are cancelled out on Array $\mathbb A$ since $\eta_1 \bm{a}(\theta_1)+\ldots+\eta_F \bm{a}(\theta_F)=\bm{0}$ holds for every snapshot. In real cases, the probability for a collection of signals to be vanishing is extremely low. We assume that the signals are non-vanishing on both Array $\mathbb A$ and Array $\mathbb B$.

In the DOA estimation using coprime arrays, one needs to estimate $\left\{ \theta_q \right\}_{q=1}^Q$ from $T_s$ snapshots of the measurements $\left\{ \bm{y}_A(t), \bm{y}_B(t)\right\}_{t=1}^{T_s}$.
The existing methods rely on the sensor-by-sensor correlations of the received signal \cite{Vaidyanathan2011}, for example, the signals on the $i_A$th sensor of Array $\mathbb A$ and the $i_B$th sensor of Array $\mathbb B$. Suppose the samples on the two sensors (omitting the additive noises) are respectively
\begin{IEEEeqnarray} {rCl} \IEEEyesnumber \IEEEyessubnumber*
y_{A}(i_A,t) &=& \sum_{q=1}^Q s_q(t) e^{j M i_A \theta_q}, \\
y_{B}(i_B,t) &=& \sum_{q=1}^Q s_q(t) e^{j N i_B \theta_q}.
\end{IEEEeqnarray}
When the $Q$ signals are statistically independent, $E\{s_p(t) s_q^*(t)\} =0$ for $p \neq q$. The correlation becomes
\begin{IEEEeqnarray} {c} 
E\left\{ y_A(i_A,t) y_B^*(i_B,t) \right\} = \sum_{q=1}^Q E | s_q(t) |^2 e^{j\left( M i_A - N i_B \right)\theta_q}.\IEEEeqnarraynumspace
\end{IEEEeqnarray}
Taking all the integer combinations $(i_A,i_B)$, $M i_A - N i_B$ traverses all the integers between $-MN$ and $MN$. The rearranged spatial autocorrelations are therefore a superposition of $Q$ sinusoids on an virtual ULA of size ${\cal O}(MN)$. A much larger correlation matrix can be constructed to resolve ${\cal O}(MN)$ signals by the subspace-based methods like MUSIC \cite{Vaidyanathan2011} \cite{Pal2011}.

However, the presence of coherent signals indicates that $E\left\{ y_A(i_A,t) y_B^*(i_B,t) \right\}$ contains cross-terms. For example, if two signals from $\theta_p$ and $\theta_q$ are coherent, the following component included in the correlation is non-zero:
\begin{IEEEeqnarray} {c}
E\{s_p(t) s_q^*(t)\} e^{j\left( M i_A \theta_p - N i_B \theta_q\right)}.
\end{IEEEeqnarray}
Since $\theta_p \neq \theta_q$, the cross-term is not corresponding to any sinusoid component on the virtual ULA. When the rearranged spatial autocorrelations are used to form a correlation matrix as in \cite{Pal2010} or \cite{Liu2015a}, the signal subspace structure is strongly contaminated, leading to a failed DOA estimation. A demonstrative example of a failed DOA estimation is given in \figurename \ref{fig:Correlation} in the simulations.

Based on the above signal model, in Section \ref{sec:FCM}, we formulate an FCM of the coprime array signal that can be adopted by the fourth-order DOA estimation. Then, a generalized spatial smoothing scheme, which is crucial for resolving coherent signals from the FCM, is introduced in Section \ref{sec:SS}.

\section{Formulation of FO cumulant matrix} \label{sec:FCM}
In this section, we begin by revising the FO cumulants of a random vector. Next, we formulate an FCM for the coprime array signal. The subspace structure of the FCM is carefully analyzed.

\subsection{FCM of signal amplitudes}
Under the assumption that the signal amplitudes are symmetrically distributed, the FCM of the signal amplitude vector $\bm{s}(t)$, denoted by $\bm{\Psi} (\bm{s})$, is well defined and given in \cite{Porat1991}
\begin{IEEEeqnarray} {rCl}  \label{eq:AmpFOC}
\bm{\Psi} (\bm{s}) &=& E \left\{ \left( \bm{s}(t) \otimes \bm{s}^*(t) \right) \left( \bm{s}(t) \otimes \bm{s}^*(t)  \right)^H \right\} \nonumber \\
&& - \> E \left\{ \bm{s}(t) \otimes \bm{s}^*(t) \right\} E \left\{ \bm{s}(t) \otimes \bm{s}^*(t) \right\}^H \nonumber \\
&& - \> E \left\{ \bm{s}(t) \bm{s}^H(t) \right\} \otimes E \left\{ \bm{s}(t) \bm{s}^H(t) \right\}^*.
\end{IEEEeqnarray}

In the $g$th coherent group, the FCM of the group signal amplitude vector $\bm{s}_g(t)$ is similarly formulated as in (\ref{eq:AmpFOC}), and is denoted by $\bm{\Psi}(\bm{s}_g)$. Since the elements in ${\bm s}_g(t)$ are linearly dependent, substituting (\ref{eq:cohcoef}) into (\ref{eq:AmpFOC}), the FCM of $\bm{s}_g(t)$ becomes
\begin{IEEEeqnarray} {c} \label{eq:AmFCM}
\bm{\Psi}(\bm{s}_g) = \left( \bm{\eta}_g \otimes \bm{\eta}_g^* \right) \Psi(\sigma_g)
\left( \bm{\eta}_g \otimes \bm{\eta}_g^* \right)^H.
\end{IEEEeqnarray}
In (\ref{eq:AmFCM}), $\bm{\eta}_g \otimes \bm{\eta}_g^*$ is a vector of length $Q_g^2$, and $\Psi(\sigma_g)$ is a scalar, detailed by
\begin{IEEEeqnarray} {rCl}
\Psi(\sigma_g) &=& E \left\{ \sigma(t)\sigma^*(t)\sigma^*(t)\sigma(t) \right\} \nonumber \\
&& \> - 2 E\left\{ \sigma(t)\sigma^*(t) \right\} E\left\{ \sigma^*(t)\sigma(t) \right\}.
\end{IEEEeqnarray}
$\Psi(\sigma_g)$ represents the FO cumulant of the source $\sigma_g(t)$. When the source follows a circularly symmetric non-Gaussian distribution, $\Psi(\sigma_g)$ is non-zero. To sum up, $\bm{\Psi}(\bm{s}_g)$ is a $Q_g^2 \times Q_g^2$ matrix with rank one, which describes the FO cumulants of the amplitude vector.

\subsection{FCM of coprime array signal}
Defining an auxiliary vector 
\begin{IEEEeqnarray} {c}
\bm{z}(t) = \bm{y}_A(t) \otimes \bm{y}_B^*(t),
\end{IEEEeqnarray}
the following matrix is the FO moments of the array signal:
\begin{IEEEeqnarray} {rCl} \label{eq:FOM}
\bm{\Gamma}_4(\bm{z}) &=& E \left\{ \bm{z}(t) \bm{z}^H(t) \right\}  \nonumber \\
 &=& E \left\{ \left( \bm{y}_{A}(t) \otimes \bm{y}_{B}^*(t) \right) \left( \bm{y}_{A}(t) \otimes \bm{y}_{B}^*(t) \right)^H \right\}.
\end{IEEEeqnarray}
The autocorrelation matrices of the array signal on Array $\mathbb A$ and Array $\mathbb B$, and the cross-correlation vector between sparse arrays are the second-order moments of the array signal, respectively formulated as
\begin{IEEEeqnarray} {rCl} \label{eq:SOM} \IEEEyesnumber \IEEEyessubnumber*
\bm{\Gamma}_2(\bm{y}_{A}) &=& E \left\{ \bm{y}_{A}(t) \bm{y}_{A}^H(t) \right\}, \\
\bm{\Gamma}_2(\bm{y}_{B}) &=& E \left\{ \bm{y}_{B}(t) \bm{y}_{B}^H(t) \right\}, \\
\bm{\mu}_2(\bm{z}) &=&  E\left\{ \bm{y}_A(t) \otimes \bm{y}_B^*(t) \right\}.
\end{IEEEeqnarray}

We now introduce a matrix consisting of the FO cumulants of the received signal. The matrix is a combination of the FO and second-order moments defined above, formulated as
\begin{IEEEeqnarray} {c} \label{eq:FOCdef}
\bm{\Phi} = {\bm \Gamma}_4(\bm{z}) - \bm{\mu}_2(\bm{z}) \bm{\mu}_2^H(\bm{z}) - \bm{\Gamma}_2(\bm{y}_{A}) \otimes \bm{\Gamma}_2^*(\bm{y}_{B}).
\end{IEEEeqnarray}
Each element of $\bm \Phi$ is a FO moment of the array signal.
Since the additive noises are statistically independent of the signal, and the FO cumulants of the Gaussian noise are identically zero \cite{Cardoso1995}, substituting (\ref{eq:Receive}), (\ref{eq:AmpFOC}), (\ref{eq:FOM}) and (\ref{eq:SOM}) into (\ref{eq:FOCdef}), we obtain
\begin{IEEEeqnarray} {rCl} \label{eq:FOC}
\bm{\Phi} &=& E \left\{ \left( \left( \bm{A} \bm{s}(t) \right) \otimes \left( \bm{B} \bm{s}(t) \right)^* \right)
   \left( \left( \bm{A} \bm{s}(t) \right) \otimes \left( \bm{B} \bm{s}(t) \right)^*\right)^H \right\} \nonumber \\
&&- E \left\{ \left( \bm{A} \bm{s}(t) \right) \otimes \left( \bm{B} \bm{s}(t) \right)^* \right\}
   E \left\{ \left( \bm{A} \bm{s}(t) \right) \otimes \left( \bm{B} \bm{s}(t) \right)^* \right\}^H \nonumber \\
&&- E\left\{ \bm{A} \bm{s}(t) \bm{s}^H(t) \bm{A}^H \right\} \otimes
   E\left\{ \bm{B} \bm{s}(t) \bm{s}^H(t) \bm{B}^H \right\}^* \nonumber \\
&=& \left(\bm{A} \otimes \bm{B}^*\right) \bm{\Psi} \left(\bm{s} \right) \left(\bm{A} \otimes \bm{B}^*\right)^H.
\end{IEEEeqnarray}

The special structure of $\bm \Phi$ informs the array configuration and the amplitude FO property. Most importantly, $\bm{A} \otimes \bm{B}^*$ is built up by the steering vectors of impinging signals, which implies the DOAs. In the remainder of this paper, ${\bm \Phi}$ is referred to as the FCM of Coarray $\mathbb{AB}$.

In practical situations, the theoretical FCM is unknown and has to be estimated. If the signal is second-order and fourth-order ergodic, the empirical estimates of the moments are
\begin{IEEEeqnarray} {rCl} \IEEEyesnumber \IEEEyessubnumber*
\hat{\bm \Gamma}_4(\bm{z}) &=& {1\over T_s} \sum_{t=1}^{T_s}  \left( \bm{y}_{A}(t) \otimes \bm{y}_{B}^*(t) \right) \left( \bm{y}_{A}(t) \otimes \bm{y}_{B}^*(t) \right)^H, \IEEEeqnarraynumspace \\
\hat{\bm \Gamma}_2(\bm{y}_{A}) &=& {1\over T_s} \sum_{t=1}^{T_s}  \bm{y}_{A}(t) \bm{y}_{A}^H(t), \\
\hat{\bm \Gamma}_2(\bm{y}_{B}) &=& {1\over T_s} \sum_{t=1}^{T_s} \bm{y}_{B}(t) \bm{y}_{B}^H(t), \\
\hat{\bm \mu}_2(\bm{z}) &=& {1\over T_s} \sum_{t=1}^{T_s} \bm{y}_{A}(t) \otimes \bm{y}_{B}^*(t).
\end{IEEEeqnarray}
The FCM of Coarray $\mathbb{AB}$ can be estimated by
\begin{IEEEeqnarray} {c}
\hat{\bm \Phi} = \hat{\bm \Gamma}_4(\bm{z}) - \hat{\bm \mu}_2(\bm{z}) \hat{\bm \mu}_2^H(\bm{z}) - \hat{\bm \Gamma}_2(\bm{y}_{A}) \otimes \hat{\bm \Gamma}_2^*(\bm{y}_{B}).
\end{IEEEeqnarray}

\subsection{Subspace structure of FCM}
In (\ref{eq:FOC}), since the cumulants of sums of independent processes are the sums of the individual cumulants \cite{Chevalier2005}, the FCM $\bm \Phi$ is the sum of FCMs of the individual coherent groups
\begin{IEEEeqnarray} {rCl} \label{eq:FocGroup} 
\bm{\Phi} &=& \sum_{g=1}^G \bm{\Phi}(g), \nonumber \\ 
\bm{\Phi}(g) &=& \left(\bm{A}(g) \otimes \bm{B}^*(g)\right) \bm{\Psi} \left(\bm{s}_g \right) \left(\bm{A}(g) \otimes \bm{B}^*(g) \right)^H, \IEEEeqnarraynumspace
\end{IEEEeqnarray}
where $\bm{\Phi}(g)$ is the FCM of the $g$th group, $\bm{A}(g)$ and $\bm{B}(g)$ are the steering vector matrices for the $g$th group on Array $\mathbb A$ and $\mathbb B$, respectively:
\begin{IEEEeqnarray} {rCl}  \IEEEyesnumber \IEEEyessubnumber*
\bm{A}(g) &=& \left[\bm{a}(\theta_{g_1}),\ldots,\bm{a}(\theta_{g_{Q_g}}) \right], \\
\bm{B}(g) &=& \left[\bm{b}(\theta_{g_1}),\ldots,\bm{b}(\theta_{g_{Q_g}}) \right].
\end{IEEEeqnarray}
When $\bm \Phi$ is eigen-decomposed, it forms a signal-subspace (with the projection operator $\bm \Pi$) spanned by the eigenvectors corresponding to the large eigenvalues, and a noise-subspace (with the projection operator $\bm \Pi^{\bot}$) spanned by the eigenvectors corresponding to the small eigenvalues. The two subspaces are orthogonal.

From (\ref{eq:FocGroup}), the signal subspace of $\bm{\Phi}(g)$ is spanned by the column vectors in the matrix $\bm{A}(g) \otimes \bm{B}^*(g)$. 
The signal subspace of $\bm{\Phi}$ is the direct sum of the signal subspaces of each $\bm{\Phi}(g)$.
Obviously, $\bm{A}(g) \otimes \bm{B}^*(g)$ includes the column vectors $\{ \bm{a}(\theta_{g_q}) \otimes \bm{b}^*(\theta_{g_q}) \}_{q=1}^{Q_g}$.
We can resort to the 4-MUSIC algorithm \cite{Porat1991} to estimate the DOAs in the following manner. 
For a signal from the DOA $\theta$, the vector $\bm{a}(\theta) \otimes \bm{b}^*(\theta)$ is in the signal subspace of $\bm \Phi$, then there exists a null at $\theta$ on the so-called null-spectrum:
\begin{IEEEeqnarray} {c}
h(\theta) = \left\| \bm{\Pi}^{\bot} \left( \bm{a}(\theta) \otimes \bm{b}^*(\theta) \right) \right\|^2. 
\end{IEEEeqnarray}
The pseudo-spectrum, or $1/h(\theta)$, is used to locate the DOAs by searching for the maxima on it.

\subsection{Impact of coherent signals on FCM}
For an independent signal, itself constitutes a group, say the $g$th group, with $Q_g=1$. As in (\ref{eq:AmFCM}), $\bm{\Psi}(\bm{s}_g)$ becomes a non-zero scalar. The contribution of the independent signal to the FCM as in (\ref{eq:FocGroup}) is
\begin{IEEEeqnarray} {c}
\bm{\Phi}(g) = \left( \bm{a}(\theta_{g_1}) \otimes \bm{b}^*(\theta_{g_1}) \right) \bm{\Psi}(\bm{s}_g)
\left( \bm{a}(\theta_{g_1}) \otimes \bm{b}^*(\theta_{g_1}) \right)^H. \IEEEeqnarraynumspace
\end{IEEEeqnarray}
Clearly, the signal subspace of $\bm{\Phi}(g)$ is spanned by the vector $\bm{a}(\theta_{g_1}) \otimes \bm{b}^*(\theta_{g_1})$. Then, a peak appears at $\theta_{g_1}$ on the 4-MUSIC pseudo-spectrum.

For a group with multiple coherent signals that $Q_g \geq 2$, $\bm{\Psi}(\bm{s}_g)$ is a $Q_g^2 \times Q_g^2$ matrix with rank one. Then, the signal subspace of $\bm{\Phi}(g)$ has only one dimension. Substituting (\ref{eq:AmFCM}) into (\ref{eq:FocGroup}), we derive
\begin{IEEEeqnarray} {c} \label{eq:CoGroup}
\bm{\Phi}(g) = \left(\bm{A}(g) \otimes \bm{B}^*(g)\right) \left( \bm{\eta}_g \otimes \bm{\eta}_g^* \right) \Psi(\sigma_g)  \nonumber \\
\cdot \left( \bm{\eta}_g \otimes \bm{\eta}_g^* \right)^H \left(\bm{A}(g) \otimes \bm{B}^*(g) \right)^H.
\end{IEEEeqnarray}
In (\ref{eq:CoGroup}), the one-dimension signal subspace of $\bm{\Phi}(g)$ is spanned by the vector 
\begin{IEEEeqnarray} {c} \label{eq:CombVec}
\left(\bm{A}(g) \otimes \bm{B}^*(g)\right) \left( \bm{\eta}_g \otimes \bm{\eta}_g^* \right).
\end{IEEEeqnarray}
The vector in (\ref{eq:CombVec}) is a linear combination of the $Q_g^2$ column vectors $\left\{ \bm{a}(\theta_{g_p}) \otimes \bm{b}^*(\theta_{g_q}) \right\}_{p,q=1}^{Q_g}$.
In the coherent signal case, the column vectors in $\bm{A}(g) \otimes \bm{B}^*(g)$ are merged. Hence, the signal subspace of $\bm{\Phi}(g)$ has only one dimension. The 4-MUSIC algorithm therefore fails to locate the DOAs of coherent signals.

\section{Generalized spatial smoothing on FCM} \label{sec:SS}
In this section, we introduce a generalized spatial smoothing scheme applied to the FCM $\bm{\Phi}$. The scheme leads to a successful estimation of the coherent signals.

\begin{figure}[!t]
\centering
\includegraphics[width=8.5cm]{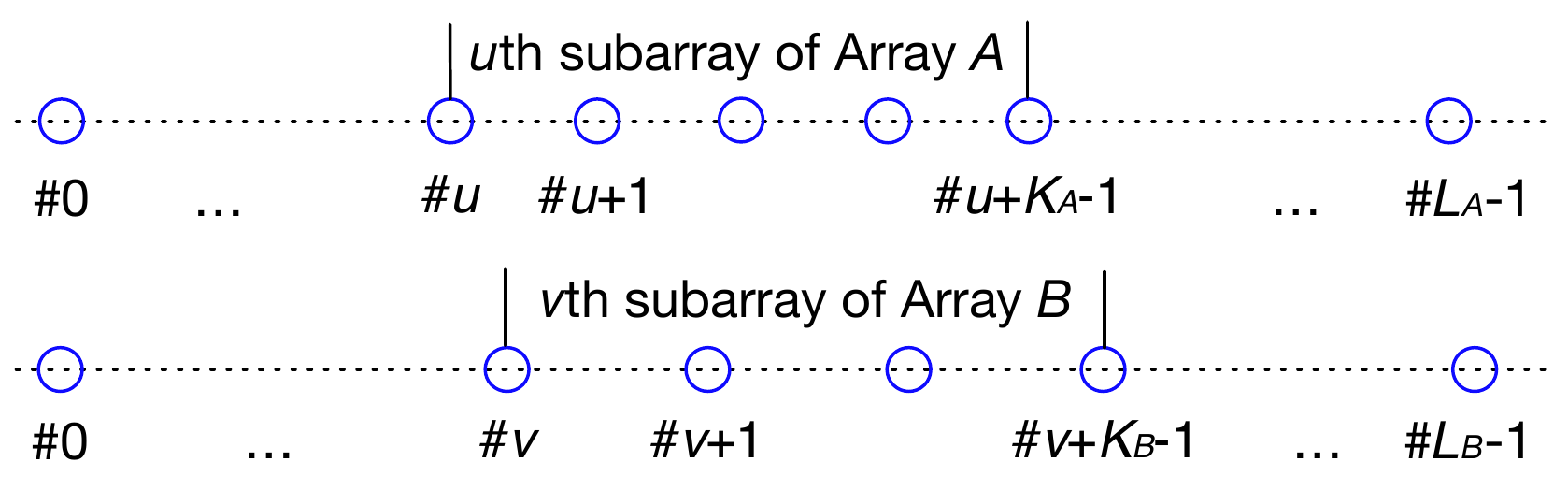} 
\caption{$(u,v)$ sub-coarrays formed by two sparse subarrays} \label{fig:SubCoarray}
\end{figure}

Since the two sparse arrays are both uniform, we can divide each of them, for example, Array $\mathbb A$ into overlapping subarrays. Every subarray contains $K_A$ continuous sensors, with sensors $\left\{0, 1, \ldots, K_A-1 \right\}$ forming the $0$th subarray, sensors $\left\{ 1, \ldots, K_A \right\}$ forming the $1$st subarray, etc. Similarly, Array $\mathbb B$ is divided into overlapping subarrays of size $K_B$. Choosing the $u$th subarray of Array $\mathbb A$ and the $v$th subarray of Array $\mathbb B$, a pair of coprime subarrays can be formed, and is denoted by the $(u,v)$ sub-coarrays. The subarray indices can be chosen from $u=0,1,\ldots,{L}_A-{K}_A$ and $v=0,1,\ldots, {L}_B-{K}_B$.
An illustration of the $(u,v)$ sub-coarrays is in \figurename \ref{fig:SubCoarray}.

For a DOA $\theta$, the partial steering vectors on the $u$th subarray of Array $\mathbb A$, and the $v$th subarray of Array $\mathbb B$ are denoted by
\begin{IEEEeqnarray} {rCl}  \IEEEyesnumber \IEEEyessubnumber*
\bm{a}_u(\theta) &=& \left[e^{jMu\theta},\ldots, e^{jM(u+{K}_A -1)\theta}  \right]^T, \\
\bm{b}_v(\theta) &=& \left[e^{jNv\theta},\ldots, e^{jN(v+{K}_B -1)\theta}  \right]^T.
\end{IEEEeqnarray}
The matrices of the collection of partial steering vectors in the $g$th group are denoted by
\begin{IEEEeqnarray} {rCl}  \IEEEyesnumber \IEEEyessubnumber*
\bm{A}_u(g) &=& \left[ \bm{a}_u(\theta_{g_1}),\ldots,\bm{a}_u(\theta_{g_{Q_g}}) \right], \\
\bm{B}_v(g) &=& \left[ \bm{b}_v(\theta_{g_1}),\ldots,\bm{b}_v(\theta_{g_{Q_g}}) \right].
\end{IEEEeqnarray}

On the $(u,v)$ sub-coarrays, an FCM is inherently defined analogous to (\ref{eq:FOC}). The size of the sub-coarray FCM is determined by the sub-coarray size $K_A,K_B$. As in (\ref{eq:FocGroup}), the sub-coarray FCM can be decomposed into the sum of contributions from each group
\begin{IEEEeqnarray} {rCl} \label{eq:subFOC}
\bm{\Phi}_{u,v} &=& \sum_{g=1}^G \bm{\Phi}_{u,v}(g), \nonumber \\
\bm{\Phi}_{u,v}(g) &=& \left(\bm{A}_u(g) \otimes \bm{B}_v^*(g)\right) \bm{\Psi} 
\left(\bm{s}_g \right) \left(\bm{A}_u(g) \otimes \bm{B}_v^*(g) \right)^H. \label{eq:groupFOC} \IEEEeqnarraynumspace
\end{IEEEeqnarray}

Comparing the $u$th subarray and the $0$th subarray on Array $\mathbb A$, and comparing the $v$th subarray and the $0$th subarray on Array $\mathbb B$, a relationship exists between the following matrices:
\begin{IEEEeqnarray} {rCl} \label{eq:Rotation} \IEEEyesnumber \IEEEyessubnumber*
\bm{A}_u(g) &=& \bm{A}_0(g) \bm{\Omega}_A^u(g), \\
\bm{B}_v(g) &=& \bm{B}_0(g) \bm{\Omega}_B^v(g).
\end{IEEEeqnarray}
In (\ref{eq:Rotation}), $\bm{\Omega}_A^u(g)$ and $\bm{\Omega}_B^v(g)$ are the $u$th and $v$th power of the following $Q_g \times Q_g$ diagonal matrices, respectively:
\begin{IEEEeqnarray} {rCl} \IEEEyesnumber \IEEEyessubnumber*
\bm{\Omega}_A(g) &=& {\rm diag} \left[ e^{jM\theta_{g_1}},\ldots, e^{jM\theta_{g_{Q_g}}}\right], \\
\bm{\Omega}_B(g) &=& {\rm diag} \left[ e^{jN\theta_{g_1}},\ldots, e^{jN\theta_{g_{Q_g}}}\right].
\end{IEEEeqnarray}
Comparing the $(u,v)$ sub-coarrays with the $(0,0)$ sub-coarrays, the following relation exists
\begin{IEEEeqnarray} {rCl}
\bm{A}_u(g) \otimes \bm{B}_v^*(g) &=& \left( \bm{A}_0(g) \bm{\Omega}_A^u(g) \right) 
\otimes \left( \bm{B}_0^*(g) \bm{\Omega}_B^{-v}(g) \right) \nonumber \\
&=& \left( \bm{A}_0(g) \otimes \bm{B}_0^*(g) \right) \bm{\Omega}^{u,-v}(g),
\end{IEEEeqnarray}
where $\bm{\Omega}^{u,-v}(g) =\bm{\Omega}_A^u(g) \otimes \bm{\Omega}_B^{-v}(g)$ is a $Q_g^2 \times Q_g^2$ diagonal matrix. 
Therefore, $\bm{\Phi}_{u,v}(g)$ can be written as
\begin{IEEEeqnarray} {l} \label{eq:focRelation}
\bm{\Phi}_{u,v}(g) = \left(\bm{A}_0(g) \otimes \bm{B}_0^*(g)\right) \nonumber \\
\quad \cdot \left[ \bm{\Omega}^{u,-v}(g) \bm{\Psi}\left(\bm{s}_g \right) \bm{\Omega}^{-u,v}(g) \right]
\left(\bm{A}_0(g) \otimes \bm{B}_0^*(g) \right)^H.
\end{IEEEeqnarray}
We observe that $\bm{\Phi}_{u,v}(g)$ and $\bm{\Phi}_{0,0}(g)$ share the same signal subspace, and are related by a `rotation' of the matrix $\bm{\Psi}(\bm{s}_g)$.

The {\it generalized spatial smoothed} FCM of coprime arrays is defined as the sum of FCMs on all the similar sub-coarrays:
\begin{IEEEeqnarray} {c} \label{eq:focSS}
\bar{\bm \Phi} = \sum_{u=0}^{{L}_A-{K}_A} \sum_{v=0}^{{L}_B-{K}_B} \bm{\Phi}_{u,v}.
\end{IEEEeqnarray}
Denote $\bar{\bm \Psi}(\bm{s}_g)$ as the smoothed FCM of the group amplitude vector $\bm{s}_g(t)$, written as
\begin{IEEEeqnarray} {c} \label{eq:AmpFocSS}
\bar{\bm \Psi}(\bm{s}_g) = \sum_{u=0}^{{L}_A-{K}_A} \sum_{v=0}^{{L}_B-{K}_B}
 \bm{\Omega}^{u,-v}(g) \bm{\Psi} \left(\bm{s}_g \right) \bm{\Omega}^{-u,v}(g).
\end{IEEEeqnarray}
Substituting (\ref{eq:subFOC}), (\ref{eq:focRelation}) and (\ref{eq:AmpFocSS}) into (\ref{eq:focSS}), the smoothed FCM $\bar{\bm \Phi}$ of the coprime arrays signal is 
\begin{IEEEeqnarray} {rCl} \label{eq:SS}
\bar{\bm{\Phi}} &=& \sum_{g=1}^G \bar{\bm{\Phi}}(g), \nonumber \\
\bar{\bm{\Phi}}(g) &=& \left(\bm{A}_0(g) \otimes \bm{B}_0^*(g)\right) \bar{\bm \Psi}(\bm{s}_g)
\left(\bm{A}_0(g) \otimes \bm{B}_0^*(g)\right)^H. \label{eq:SSgroup} \IEEEeqnarraynumspace
\end{IEEEeqnarray}
In (\ref{eq:AmpFocSS}), $\bar{\bm \Psi}({\bm s}_g)$ is rank-enhanced from ${\bm \Psi}({\bm s}_g)$. Consequently in (\ref{eq:SS}), the smoothed FCM $\bar{\bm \Phi}$ is rank-enhanced. The effect of the generalized spatial smoothing scheme on the FCM is analogous to the spatial smoothing scheme on the correlation matrix of a ULA \cite{Shan1985}.

An important theorem is in place here. We show that with some restrictions, the vectors $\left\{ \bm{a}_0(\theta_q) \otimes \bm{b}_0^*(\theta_q) \right\}_{q=1}^{Q}$ for all the signals are in the signal subspace of $\bar{\bm \Phi}$.

\newtheorem{theorem}{Theorem}
\begin{theorem} \label{th:rank}
In the $g$th group, if $ L_A -  K_A +1 \geq Q_g$ and $ L_B -  K_B +1 \geq Q_g$, the vector $\bm{a}_0(\theta_{g_q}) \otimes \bm{b}_0^*(\theta_{g_q})$ for any one of the DOAs $\{\theta_{g_q}\}_{q=1}^{Q_g}$ is in the signal subspace of $\bar{\bm \Phi}(g)$.
\end{theorem}

\begin{IEEEproof}
See Appendix.
\end{IEEEproof} 

\newtheorem{corollary}{Corollary}
\begin{corollary} \label{cor:subspace}
Let $Q_{max}={\rm max} (Q_1,\ldots,Q_G)$. If $ L_A -  K_A +1 \geq Q_{max}$, $ L_B -  K_B +1 \geq Q_{max}$, then the vector $\bm{a}_0(\theta_q) \otimes \bm{b}_0^*(\theta_q)$ for any one of the DOAs $\{\theta_q\}_{q=1}^Q$ is in the signal subspace of the smoothed FCM $\bar{\bm \Phi}$.
\end{corollary}

\begin{IEEEproof}
Follows Theorem \ref{th:rank} and that $\bar{\bm{\Phi}} = \sum_{g=1}^G \bar{\bm{\Phi}}(g)$, the signal subspace of $\bar{\bm \Phi}$ is the direct sum of all the signal subspaces of $\bar{\bm \Phi}(g), g=1,\ldots,G$.
\end{IEEEproof} 

Corollary \ref{cor:subspace} indicates that, if the numbers of overlapping subarrays on both sparse arrays are no less than the largest number of the coherent signals, the vectors $\left\{ \bm{a}_0(\theta_q) \otimes \bm{b}_0^*(\theta_q) \right\}_{q=1}^{Q}$ for both the independent and coherent signals are in the signal subspace of $\bar{\bm{\Phi}}$. Then one can eigen-decompose $\bar{\bm{\Phi}}$ to acquire a noise subspace with the projection operator $\bar{\bm \Pi}^{\bot}$. The vector $\bm{a}_0(\theta) \otimes \bm{b}_0^*(\theta)$ for any signal with a DOA $\theta$ is orthogonal to the noise subspace. From the smoothed FCM, the null-spectrum produced by 4-MUSIC is defined as
\begin{IEEEeqnarray} {c}
\bar{h}(\theta) = \left\| \bar{\bm \Pi}^{\bot} \left( \bm{a}_0(\theta) \otimes \bm{b}_0^*(\theta) \right) \right\|^2. 
\end{IEEEeqnarray}
On the pseudo-spectrum $1 / \bar{h}(\theta)$, both the independent and coherent signals create peaks at their respective directions.

Remark: The sub-coarray FCM $\bm{\Phi}_{u,v}$ in (\ref{eq:subFOC}) is a principle sub-matrix of the full-coarray FCM $\bm{\Phi}$. The indices of columns (and rows) of the principle sub-matrix $\bm{\Phi}_{u,v}$ in $\bm{\Phi}$ are
\begin{IEEEeqnarray} {c} \label{eq:PIndix}
(u : u+{K}_A-1) \cdot {L}_B+(v : v+{K}_B - 1).
\end{IEEEeqnarray}
The generalized smoothing process can be accomplished by summing all the proper principle sub-matrices with the indices in (\ref{eq:PIndix}) from $\bm{\Phi}$.

The generalized spatial smoothing scheme is obviously at the expense of a reduced effective array aperture. In fact, the size of the FCM $\bm \Phi$ is $L_A  L_B \times  L_A  L_B$, while the smoothed FCM $\bar{\bm \Phi}$ is only $K_A  K_B \times  K_A  K_B$.

\section{Removing false peaks} \label{sec:FalsePeak}
On the pseudo-spectrum from the smoothed FCM, some false peaks occasionally arise at the directions where none of the true signals resides. An example of the false peaks is in \figurename \ref{fig:3Sub} in the simulations. In this section, we explain the false peak phenomenon, and provide a technique to remove them.

\subsection{Explanation of the false peaks}
When the generalized spatial smoothing scheme is applied to enhance the rank of $\bar{\bm \Psi}(\bm{s}_g)$ in (\ref{eq:SSgroup}), not only the vector $\bm{a}_0(\theta_{g_q}) \otimes \bm{b}_0^*(\theta_{g_q})$ for the signal DOAs $\{\theta_{g_q} \}_{q=1}^{Q_g}$, but also the cross-terms $\left\{ \bm{a}_0(\theta_{g_p}) \otimes \bm{b}_0^*(\theta_{g_q}) \right\}_{p \neq q}$ appear in the signal subspace of $\bar{\bm \Phi}(g)$. These cross-terms are not corresponding to any signal component, and should not create peaks on the pseudo-spectrum. However, the steering vectors $\bm{a}_0(\theta)$ and $\bm{b}_0(\theta)$ are ambiguous due to the sparsity of Array $\mathbb A$ and $\mathbb B$. We show that a false peak may appear as a result of the direction ambiguity.
Because two independent signals will not create such cross-terms, in the remainder of this section, the discussion is limited to a single coherent group. For simplicity, we omit the group index $g$.

\begin{figure}[!t]
\centering
\includegraphics[width=8.6cm]{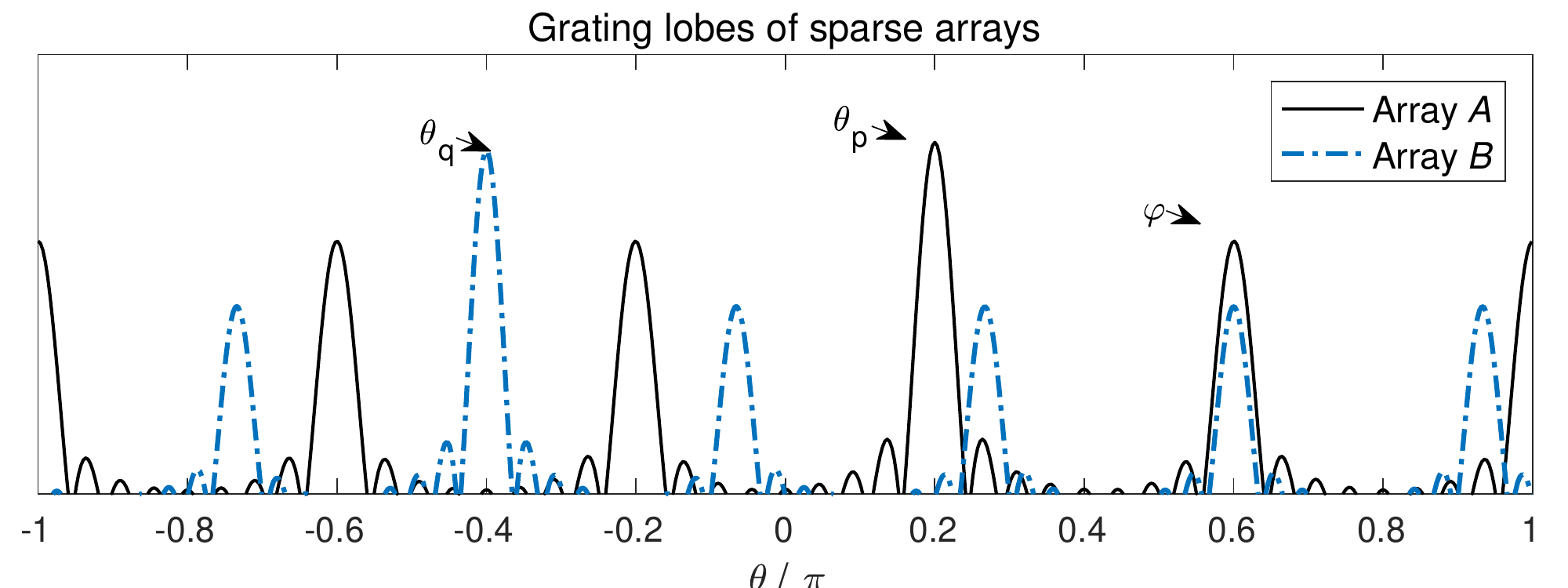}
\caption{An illustration of the beam patterns on Array $\mathbb{A}$ and $\mathbb B$. The solid line represents the grating lobe of Array $\mathbb A$, the dotted line represents the grating lobe of Array $\mathbb B$. The true DOAs $\theta_p$ and $\theta_q$ are marked by the highest lobes, while the lower lobes are the grating lobes. Theoretically, the lobes of each array should have the equal height.} \label{fig:FalsePeak}
\end{figure}

Since Array $\mathbb A$ is $M$-sparse and Array $\mathbb B$ is $N$-sparse, the steering vectors are ambiguous. In particular, for $\theta_p$ on Array $\mathbb A$ and $\theta_q$ on Array $\mathbb B$,
\begin{IEEEeqnarray} {rCl} \IEEEyesnumber \IEEEyessubnumber*
\bm{a}_0(\theta_{p}) &=& \bm{a}_0 \left(\theta_{p} + {2\pi m / M} \right),\\ 
\bm{b}_0(\theta_{q}) &=& \bm{b}_0 \left(\theta_{q} + {2\pi n / N} \right),
\end{IEEEeqnarray}
where $m,n$ are arbitrary integers. The ambiguity can be illustrated more clearly by the array beam patterns. In  \figurename ~\ref{fig:FalsePeak}, since Array $\mathbb A$ and $\mathbb B$ are both sparse and uniform, their beam patterns have multiple grating lobes.

If it happens that a grating lobe of the Array $\mathbb A$ beam pattern and a grating lobe of the Array $\mathbb B$ beam pattern overlap at the direction $\varphi$, i.e. for a pair of non-zero integers $(m,n)$,
\begin{IEEEeqnarray} {rCl} \label{eq:AmbiAB}  \IEEEyesnumber \IEEEyessubnumber* 
\varphi &=& \theta_p + 2 \pi m/M, \\
\varphi &=& \theta_q + 2 \pi n/N,
\end{IEEEeqnarray}
the cross-term of the steering vectors becomes
\begin{IEEEeqnarray} {c} \label{eq:AmbiEq}
\bm{a}_0(\theta_{p}) \otimes \bm{b}_0^*(\theta_{q}) = \bm{a}_0(\varphi) \otimes \bm{b}_0^*(\varphi).
\end{IEEEeqnarray}
When the smoothed FCM $\bar{\bm \Phi}_{AB}$ of Coarray $\mathbb{AB}$ is eigen-decomposed with the noise subspace projection operator $\bar{\bm \Pi}_{AB}$, the null-spectrum at $\varphi$ becomes
\begin{IEEEeqnarray} {rCl}
\bar{h}_{AB}(\varphi) &=& \left\| \bar{\bm \Pi}_{AB}^{\bot} \left( \bm{a}_0(\varphi) \otimes \bm{b}_0^*(\varphi) \right) \right\|^2 \nonumber \\
&=& \left\| \bar{\bm \Pi}_{AB}^{\bot} \left( \bm{a}_0(\theta_{p}) \otimes \bm{b}_0^*(\theta_{q}) \right) \right\|^2 = 0,
\end{IEEEeqnarray}
since the cross-term $\bm{a}_0(\theta_{p}) \otimes \bm{b}_0^*(\theta_{q})$ is in the signal subspace of $\bar{\bm \Phi}_{AB}$. A false peak at $\varphi$ will appear on the pseudo-spectrum.

One remark is in place here. In (\ref{eq:AmbiAB}), the direction $\varphi$ needs not to be strictly equal to $\theta_p+2\pi m/M$ or $\theta_q+2\pi n/N$. Once a grating lobe of the beam pattern $\bm{a}_0(\theta_p)$ and a grating lobe of the beam pattern $\bm{b}_0(\theta_q)$ overlap around $\varphi$, a false peak still appears.

Furthermore, the grating lobe beamwidths of the  beam patterns of $\bm{a}_0(\theta_p)$ and $\bm{b}_0(\theta_q)$ are $2\pi / M K_A$ and $2\pi / NK_B$, respectively. If $K_A \geq N$ and $ K_B \geq M$, the beamwidths of the grating lobes are narrower than $2\pi / MN$. From the observation that $M,N$ are coprime numbers, the grating lobes of the beam patterns $\bm{a}_0(\theta_p)$ on Array $\mathbb A$ and $\bm{b}_0(\theta_q)$ on Array $\mathbb B$ may overlap at one direction within $[ -\pi, \pi ]$ at most. Which means the cross-term $\bm{a}_0(\theta_{p}) \otimes \bm{b}_0^*(\theta_{q})$ may produce at most one false peak.

\subsection{Supplementary sparse array}
Suppose that a supplementary sparse array, namely Array $\mathbb C$, is deployed with its sensors positioned at
\begin{IEEEeqnarray} {c} \label{eq:3rd}
P_C = \left\{ R i_C d ~|~ i_C =0,1, \ldots, L_C-1 \right\}.
\end{IEEEeqnarray}
In (\ref{eq:3rd}), $R$ is an integer, which is respectively coprime to $M$ and $N$, and $ L_C$ is the number of sensors in Array $\mathbb C$. Now, Array $\mathbb A$ and Array $\mathbb C$ can form a new pair of coprime arrays, denoted by Coarray $\mathbb{AC}$. An FCM ${\bm \Phi}_{AC}$ can be derived for the array signal. Dividing Array $\mathbb C$ into identical subarrays of size $K_C$, we can perform the generalized spatial smoothing scheme on ${\bm \Phi}_{AC}$ to obtain the smoothed FCM $\bar{\bm \Phi}_{AC}$. Denote the noise subspace projection operator by $\bar{\bm \Pi}^{\bot}_{AC}$, the null-spectrum of Coarray $\mathbb{AC}$ is
\begin{IEEEeqnarray} {c}
{\bar h}_{AC}(\theta) = \left\| \bar{\bm \Pi}_{AC}^{\bot} \left( \bm{a}_0(\theta) \otimes \bm{c}_0^*(\theta) \right) \right\|^2,
\end{IEEEeqnarray}
where $\bm{c}_0(\theta) = \left[ e^{jR 0 \theta},\ldots, e^{jR({K}_C -1)\theta} \right]^T$. 

If $K_A \geq R$ and $K_C \geq M$ both holds, on the pseudo-spectrum $1/\bar{h}_{AC}(\theta)$ derived from Coarray $\mathbb{AC}$, the cross-term $\bm{a}_0(\theta_{p}) \otimes \bm{c}_0^*(\theta_{q})$ may produce at most one false peak, denoted by $\varphi'$. The false peak is induced by the following cross-term:
\begin{IEEEeqnarray} {c}
\bm{a}_0(\theta_{p}) \otimes \bm{c}_0^*(\theta_{q}) = \bm{a}_0(\varphi') \otimes \bm{c}_0^*(\varphi').
\end{IEEEeqnarray}
We can derive an important theorem for the position of the false peaks.

\begin{theorem} \label{th:false}
From the same pair of coherent signals $\theta_p,\theta_q$, if the cross-term $\bm{a}_0(\theta_{p}) \otimes \bm{b}_0^*(\theta_{q})$ produces a false peak at $\varphi$ on the pseudo-spectrum of Coarray $\mathbb{AB}$, and the cross-term $\bm{a}_0(\theta_{p}) \otimes \bm{c}_0^*(\theta_{q})$ produces a false peak at $\varphi'$ on the pseudo-spectrum of Coarray $\mathbb{AC}$, then $\varphi \neq \varphi'$.
\end{theorem}

\begin{IEEEproof} 
The proof is by contradiction. On Coarray $\mathbb{AC}$, the direction ambiguity indicates the following relations for a pair of non-zero integers $(m,r)$:
\begin{IEEEeqnarray} {c} \label{eq:AmbiAC}
\varphi' = \theta_p + 2 \pi m/M, \qquad \varphi' = \theta_q + 2 \pi r/R,
\end{IEEEeqnarray}
If $\varphi=\varphi'$, combining (\ref{eq:AmbiAB}) and (\ref{eq:AmbiAC}), we can deduce that $n/N= r/R$. Since $N$ and $R$ are coprime numbers, $(m,n,r)$ are within a range such that $\varphi, \varphi'$ are between $[-\pi,\pi]$, the equation holds only when $n=r=0$, which is contradictory to the non-zero assumption on the integers $n,r$.
\end{IEEEproof} 

Theorem \ref{th:false} indicates that, on two different coprime array configurations, the false peaks created by the same pair of coherent signals do not overlap on the pseudo-spectrum. This property can be used for removing the false peaks.

\subsection{Removing false peaks by combined spectrum}
To remove the false peaks induced by the direction ambiguity, we use the property that the false peaks do not overlap. From the three sparse arrays $\mathbb A$, $\mathbb B$ and $\mathbb C$, any two sparse arrays constitutes a pair of coprime arrays. Therefore, we may derive three null-spectrums: $\bar{h}_{AB}(\theta)$ from Coarray $\mathbb{AB}$, $\bar{h}_{AC}(\theta)$ from Coarray $\mathbb{AC}$ and $\bar{h}_{BC}(\theta)$ from Coarray $\mathbb{BC}$.
A combined null-spectrum can be generated from the individual null-spectrums:
\begin{IEEEeqnarray} {c} \label{eq:Combine}
\bar{h}_{ABC}(\theta) = \bar{h}_{AB}(\theta)+\bar{h}_{AC}(\theta)+\bar{h}_{BC}(\theta).
\end{IEEEeqnarray}

A necessary condition for the existence of a peak at $\theta$ on $1/ \bar{h}_{ABC}(\theta)$ is that, $\theta$ is corresponding to a null at any one of the three null-spectrums. 
In fact, if $\theta$ is the DOA of a true signal, there always exists a null at $\theta$ on any one of the three null-spectrums, $\bar{h}_{AB}(\theta)=\bar{h}_{AC}(\theta)=\bar{h}_{BC}(\theta)=0$. However, from Theorem \ref{th:false}, the false peak positions are different. 
Consequently, the false peaks are removed on the combined pseudo-spectrum $\theta$ on $1/ \bar{h}_{ABC}(\theta)$. 

\section{Simulation Results} \label{sec:Simulation}
In this section, we present some simulations that demonstrate the DOA estimation of coherent signals using the generalized spatial smoothing scheme. We also exhibit the false peak phenomenon, and the removal of false peaks on the combined spectrum.

In the first simulation, a pair of coprime arrays is used to verify the generalized spatial smoothing scheme. Array $\mathbb A$ ($M=6$) has $L_A = 9$ sensors, and Array $\mathbb B$ ($N=5$) has $L_B = 10$ sensors. The unit inter-sensor spacing is half-wavelength. We consider ten narrowband signals impinge on the coprime arrays, in which four of them are independent signals, the other six signals are divided into three coherent groups, with two signals in each group. The signal types are either QPSK or quaternary QAM, the powers of all sources are equal, and the coefficients along all the propagation paths have the equal amplitude, but with random complex phases. The noise is additive complex Gaussian, the SNR on the sensors is set to be 5dB. In the estimation of the FCM $\bm \Phi$ in (\ref{eq:FOC}), 2000 snapshots are used. In the generalized spatial smoothing scheme, Array $\mathbb A$ is divided into subarrays of size $K_A=6$, Array $\mathbb B$ is divided into subarrays of size $K_B=7$.
\begin{figure}[!t]
\centering
\subfloat[Traditional approach in \cite{Pal2011}]{\includegraphics[width=9cm]{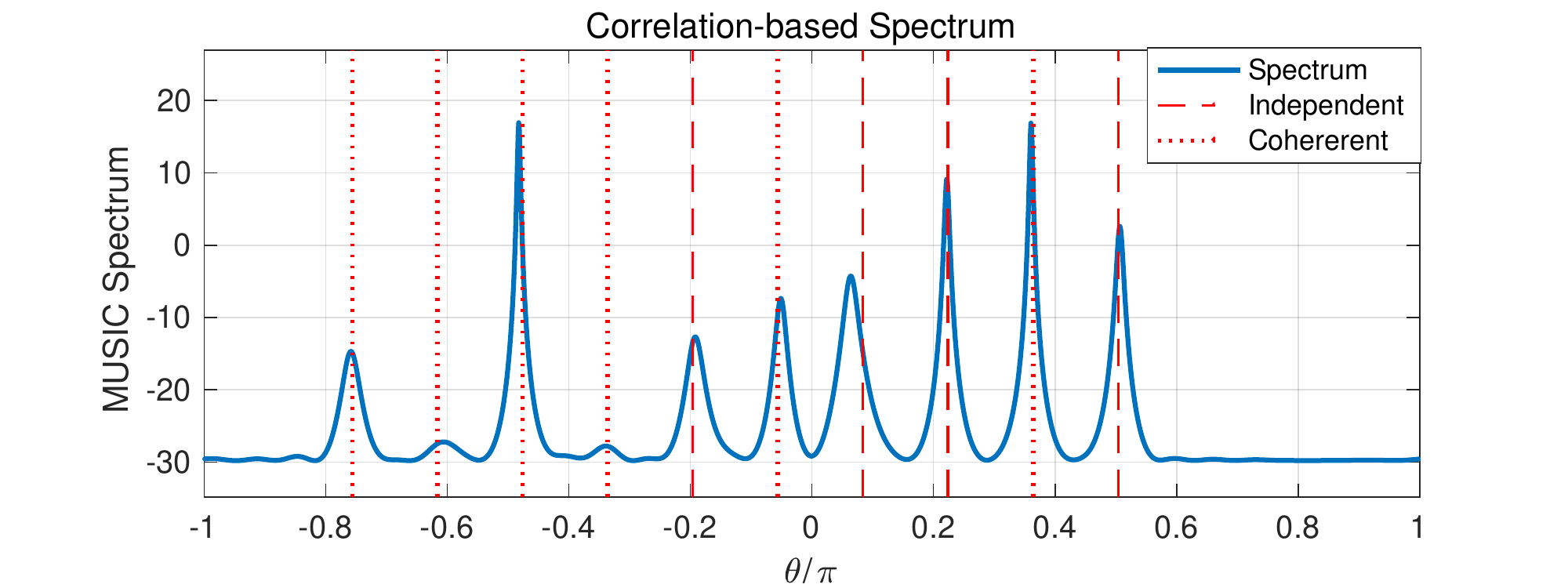} \label{fig:Correlation}}
\hfill
\subfloat[4-MUSIC with FCM $\bm \Phi$]{\includegraphics[width=9cm]{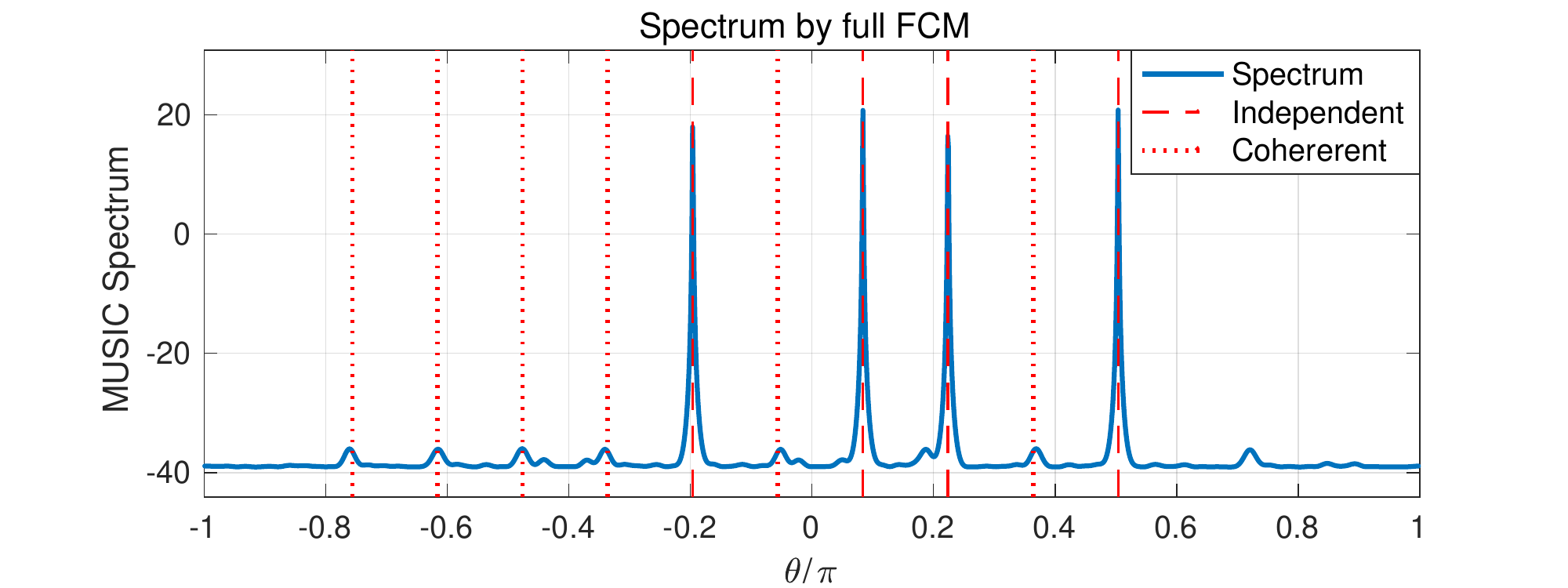} \label{fig:NoSS}}
\hfill
\subfloat[4-MUSIC with smoothed FCM $\bar{\bm \Phi}$]{\includegraphics[width=9cm]{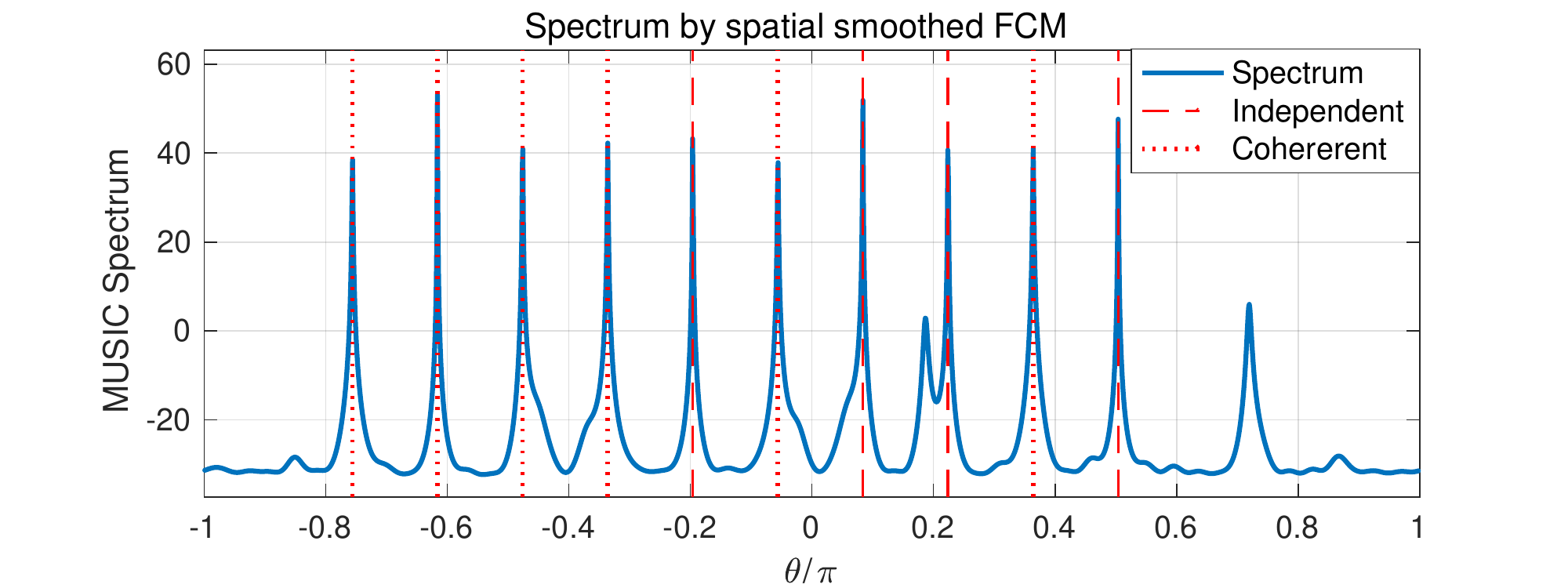} \label{fig:SS}}
\caption{The pseudo-spectrums produced by 4-MUSIC are the blue curves. The true DOA of independent signals are marked by dashed vertical lines, and the true DOA of coherent signals are marked by dotted vertical lines.}
\label{fig:SScompare}
\end{figure}

\figurename \ref{fig:Correlation} is the spectrum derived from the method in \cite{Pal2011}, which is most widely used on coprime arrays currently. The cross-terms induced by the coherent signals contaminate the structure of the signal subspace, the DOA estimation consequently fails.
In \figurename \ref{fig:NoSS}, the 4-MUSIC algorithm is applied to the FCM $\bm \Phi$. We can see that the four peaks corresponding to the independent signals appear, but the DOA of coherent signals are not resolved. On the contrary, if the generalized spatial smoothing scheme is used on the FCM, the peaks for both independent and coherent signals are clearly present in \figurename \ref{fig:SS}. It is exemplified that the DOA of coherent signals can be estimated from the smoothed FCM. 

\begin{figure}[!t]
\centering
\subfloat[Pseudo-spectrum of Coarray $\mathbb{AB}$]{\includegraphics[width=9cm]{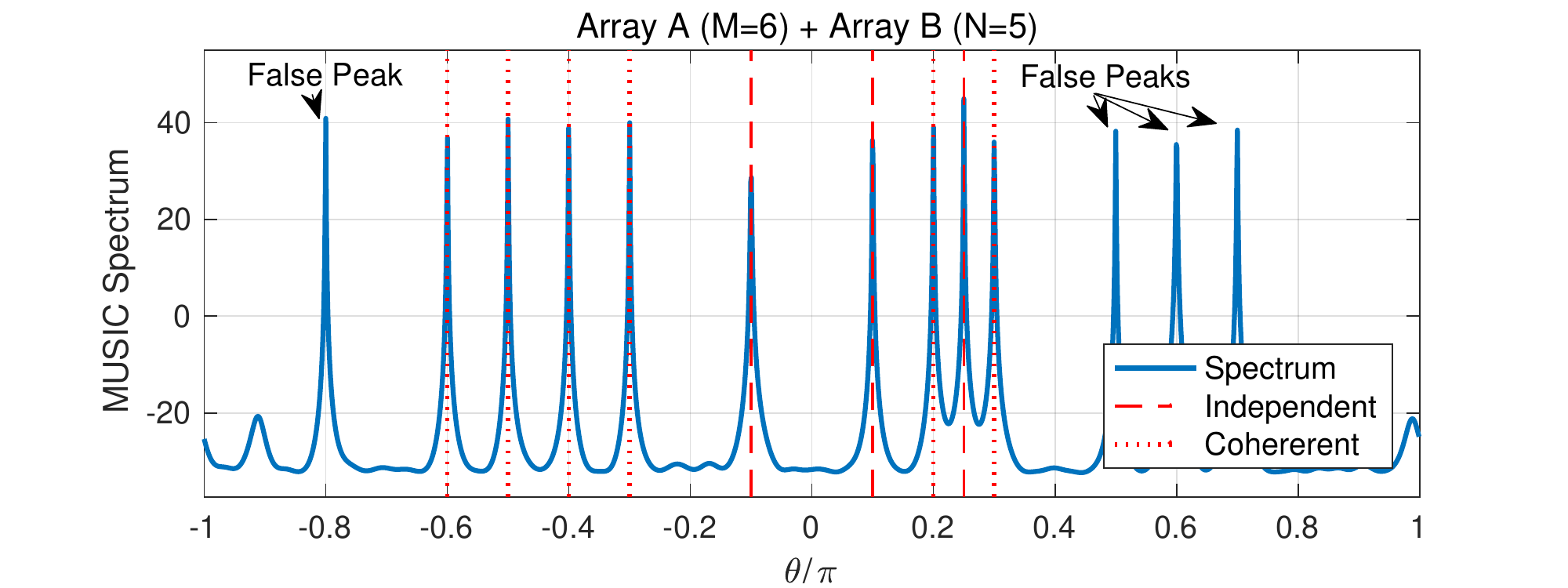} \label{fig:C56}}
\hfil
\subfloat[Pseudo-spectrum of Coarray $\mathbb{AC}$]{\includegraphics[width=9cm]{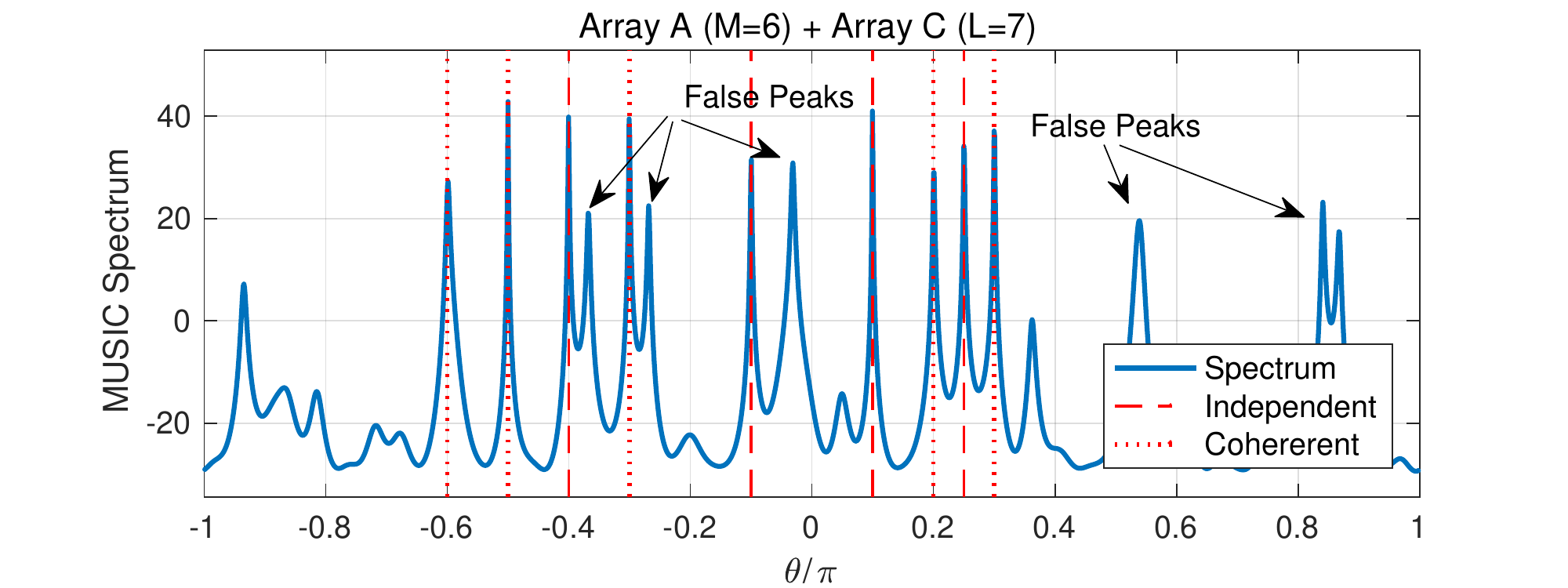} \label{fig:C67}}
\hfil
\subfloat[Pseudo-spectrum combining three null-spectrums]{\includegraphics[width=9cm]{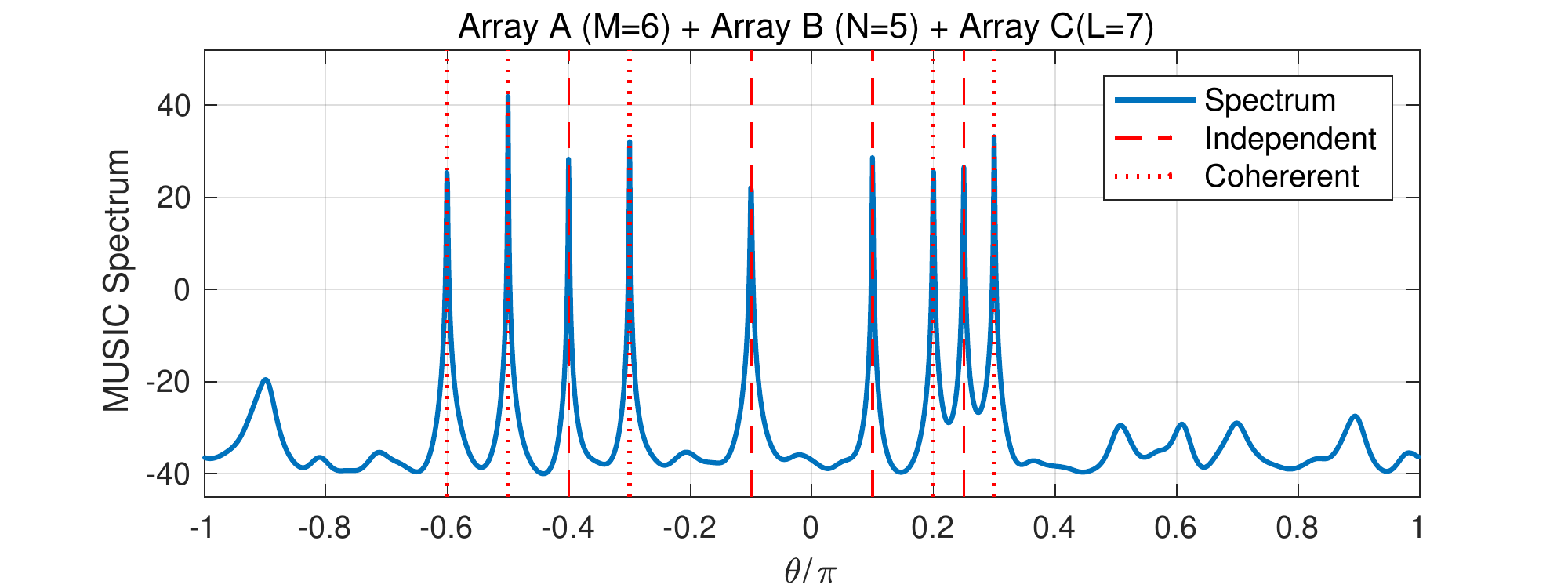} \label{fig:C567}}
\caption{The pseudo-spectrums using the smoothed FCM of two sparse arrays have false peaks as in (a) and (b). Combining the null-spectrums of the three pairs of coprime arrays following (\ref{eq:Combine}) can remove the false peaks as in (c).} \label{fig:3Sub}
\end{figure}

In the second simulation, there exist a coherent group with three signals at the normalized DOAs $\left[-0.6\pi,0.2\pi,0.3\pi\right]$, another coherent group with two signals at $\left[ -0.5\pi,-0.3\pi \right]$, and four other independent signals at $\left[ -0.4\pi,-0.1\pi,0.1\pi,0.25\pi \right]$ in the environment. The signal types, source amplitudes and propagation coefficients are set up as in the first simulation. We first derive the pseudo-spectrum using the smoothed FCM of Coarray $\mathbb{AB}$ in \figurename \ref{fig:C56}. Both independent and coherent signals are estimated on the spectrum. But there also exists multiple false peaks, which seriously affect the extraction of the true signals.

A supplementary sparse array, namely Array $\mathbb C$, with sparsity $R=7$ and $L_C=8$ sensors is deployed, we can construct a new pair of coprime arrays with Array $\mathbb A$ and Array $\mathbb C$. Let Array $\mathbb C$ be divided into subarrays of size $K_C=5$, the pseudo-spectrum using the smoothed FCM of Coarray $\mathbb{AC}$ is displayed in \figurename \ref{fig:C67}. We can see that the peaks corresponding to the true signals are still there, while the directions of the false peaks are different from that in \figurename \ref{fig:C56}.

Combining the three null-spectrum using (\ref{eq:Combine}), the pseudo-spectrum is displayed in \figurename \ref{fig:C567}. We can see that all the signals are found on the spectrum, while the false peaks are removed. Thus, we have shown the effectiveness of removing false peaks by the combined spectrum.

\section{Conclusion} \label{sec:Conclusion}
In this paper, the problem of direction-of-arrival (DOA) estimation of coherent signals on passive coprime arrays is investigated. We utilize the fourth-order cumulants to explore more information about the received signal. Formulating a fourth-order cumulant matrix (FCM) for the signal on a pair of coprime arrays, a new estimation scheme based on the fourth-order MUSIC algorithm is developed.

The special structure of the FCM is combined with the array configuration to resolve the coherent signals.
Using the property that the individual sparse arrays are uniform, on either of the sparse arrays, a series of overlapping identical subarrays can be extracted. Then, take individually one subarray from each of the sparse arrays, a pair of coprime subarrays is constructed. We revealed that the FCMs of any two similar pairs of coprime subarrays share the same structure. Analogous to the spatial smoothing scheme applied to the correlation matrix on a uniform linear array, we propose a generalized spatial smoothing scheme applied to the FCM. The scheme yields a smoothed FCM with rank-enhancement. The DOAs of both the independent and coherent signals can be estimated using the smoothed FCM.

To remove the false peaks induced by the direction ambiguity, we use a supplementary sparse array for assistance. On the combined spectrum aided by the supplementary array, the false peaks are removed while the true peaks remain. Simulation examples are given to validate the effectiveness of our approach.

\appendix[Proof of Theorem \ref{th:rank}]
Substituting (\ref{eq:AmFCM}) into (\ref{eq:AmpFocSS}), we derive
\begin{IEEEeqnarray} {c} \label{eq:RankSS}
\bar{\bm \Psi}(\bm{s}_g) = \sum_{u=0}^{{L}_A-{K}_A} \sum_{v=0}^{{L}_B-{K}_B}
 \bm{\Omega}^{u,-v}(g)  \left( \bm{\eta}_g \otimes \bm{\eta}_g^* \right) \nonumber \\
\cdot \Psi(\sigma_g) \left( \bm{\eta}_g \otimes \bm{\eta}_g^* \right)^H \bm{\Omega}^{-u,v}(g).
\end{IEEEeqnarray}
Since $\Psi(\sigma_g)$ is a non-zero scalar, we can omit it without affecting the rank of $\bar{\bm \Psi}(\bm{s}_g)$. The following discussion is limited to the $g$th group. For simplicity, the group index $g$ is omitted.

Equation (\ref{eq:RankSS}) can be written as
\begin{IEEEeqnarray} {rCl} \label{eq:sum}
\bar{\bm \Psi}(\bm{s}) &=& \sum_{u=0}^{{L}_A-{K}_A} \sum_{v=0}^{{L}_B-{K}_B}
 \bm{\Omega}^{u,-v}  \left( \bm{\eta} \otimes \bm{\eta}^* \right) \left( \bm{\eta} \otimes \bm{\eta}^* \right)^H  \bm{\Omega}^{-u,v} \nonumber \\
&=& \sum_{u=0}^{{L}_A-{K}_A} \sum_{v=0}^{{L}_B-{K}_B} {\bm \zeta}_{u,v} {\bm \zeta}_{u,v}^H,
\end{IEEEeqnarray}
where ${\bm \zeta}_{u,v}$ is a $Q^2 \times 1$ vector, written as
\begin{IEEEeqnarray} {c}
{\bm \zeta}_{u,v} = \bm{\Omega}^{u,-v} \left( \bm{\eta} \otimes \bm{\eta}^* \right) 
= \left( \bm{\Omega}_A^{u} \bm{\eta} \right) \otimes  \left( \bm{\Omega}_B^{v} \bm{\eta} \right)^*.
\end{IEEEeqnarray}
The summation in (\ref{eq:sum}) can be written in the vector inner-product form
\begin{IEEEeqnarray} {c} \label{eq:PsiW}
\bar{\bm \Psi}(\bm{s}) = \sum_{u=0}^{{L}_A-{K}_A} \sum_{v=0}^{{L}_B-{K}_B} {\bm \zeta}_{u,v} {\bm \zeta}_{u,v}^H
= {\bf W} {\bf W}^H,
\end{IEEEeqnarray}
where $\bf W$ is a matrix with $Q^2$ column vectors
\begin{IEEEeqnarray} {rCl} \label{eq:Wmat}
{\bf W} &=& \left[ {\bm \zeta}_{0,0},{\bm \zeta}_{0,1},\ldots,{\bm \zeta}_{0,{L}_B-{K}_B},\ldots,{\bm \zeta}_{{L}_A-{K}_A,{L}_B-{K}_B} \right] \nonumber \\
&=& \left[ \bm{\Omega}_A^{0}\bm{\eta}, \ldots, \bm{\Omega}_A^{{L}_A-{K}_A} \bm{\eta} \right]
\otimes 
\left[ \bm{\Omega}_B^{0}\bm{\eta}, \ldots, \bm{\Omega}_B^{{L}_B-{K}_B} \bm{\eta} \right]^* \nonumber \\
&=& {\bf W}_A \otimes {\bf W}_B^*. \IEEEeqnarraynumspace
\end{IEEEeqnarray}

In (\ref{eq:Wmat}), ${\bf W}_A$ and ${\bf W}_B$ are both $Q\times Q$ matrices. Each of them can also be written as the multiplication of a diagonal matrix and a Vandermonde matrix:
\begin{IEEEeqnarray} {rCcCl} \label{eq:Van} \IEEEyesnumber \IEEEyessubnumber*
{\bf W}_A &=& \left[ \bm{\Omega}_A^{0}\bm{\eta}, \ldots, \bm{\Omega}_A^{{L}_A-{K}_A} \bm{\eta} \right]
&=& {\rm diag}(\bm{\eta}) \cdot {\bf V}_A, \label{eq:VanA} \\
{\bf W}_B &=& \left[ \bm{\Omega}_B^{0}\bm{\eta}, \ldots, \bm{\Omega}_B^{{L}_B-{K}_B} \bm{\eta} \right] 
&=& {\rm diag}(\bm{\eta}) \cdot {\bf V}_B, \label{eq:VanB} \IEEEeqnarraynumspace
\end{IEEEeqnarray}
where the Vandermonde matrices ${\bf V}_A$ and ${\bf V}_B$ are
\begin{IEEEeqnarray} {rCl} \IEEEyesnumber \IEEEyessubnumber*
{\bf V}_A &=& \begin{bmatrix}
e^{jM0 \theta_1} & \ldots & e^{jM({L}_A-{K}_A) \theta_1} \\
\vdots & \ddots & \vdots \\
e^{jM0 \theta_Q} & \ldots & e^{jM({L}_A-{K}_A) \theta_Q}
\end{bmatrix}, \\
{\bf V}_B &=& \begin{bmatrix}
e^{jN0 \theta_1} & \ldots & e^{jN({L}_B-{K}_B) \theta_1} \\
\vdots & \ddots & \vdots \\
e^{jN0 \theta_Q} & \ldots & e^{jN({L}_B-{K}_B) \theta_Q}
\end{bmatrix}.
\end{IEEEeqnarray}

Substituting (\ref{eq:PsiW}) and (\ref{eq:Wmat}) into (\ref{eq:SSgroup}) we obtain
\begin{IEEEeqnarray} {rCl} \label{eq:PhiW} 
\bar{\bm{\Phi}} &=& \left(\bm{A}_0 \otimes \bm{B}_0^*\right) {\bf W} {\bf W}^H \left(\bm{A}_0 \otimes \bm{B}_0^*\right)^H \nonumber \\
&=& \left( \left( \bm{A}_0 {\bf W}_A \right) \otimes \left( \bm{B}_0^* {\bf W}_B^* \right) \right)
\left( \left( \bm{A}_0 {\bf W}_A \right) \otimes \left( \bm{B}_0^* {\bf W}_B^* \right) \right)^H \nonumber \\
&=& {\bm \Upsilon} {\bm \Upsilon}^H,
\IEEEeqnarraynumspace
\end{IEEEeqnarray}
where
\begin{IEEEeqnarray} {rCl} \label{eq:PhiKro} 
\bm{\Upsilon} &=& \left( \bm{A}_0 {\bf W}_A \right) \otimes \left( \bm{B}_0^* {\bf W}_B^* \right) \nonumber \\
&=& \left( \bm{A}_0 \cdot {\rm diag}(\bm{\eta}) \cdot {\bf V}_A \right) \otimes 
\left( \bm{B}_0^* \cdot {\rm diag}(\bm{\eta}^*) \cdot {\bf V}_B^* \right).
\end{IEEEeqnarray}

In (\ref{eq:PhiW}), the structure of the signal subspace of $\bar{\bm{\Phi}}$ is determined by the the rank of $\bf W$, and the column vectors in $\bm{A}_0 \otimes \bm{B}_0^*$. From (\ref{eq:Wmat}), ${\rm rank}({\bf W}) = {\rm rank}({\bf W}_A) \cdot {\rm rank}({\bf W}_B)$. In order to determine the rank of $\bf W$, the ranks of ${\bf W}_A$ and ${\bf W}_B$ are carefully discussed in the following two cases.

{\it Case I}: The DOAs are non-ambiguous on the individual sparse arrays, which means the steering vectors on Array $\mathbb A$ $\{\bm{a}_0(\theta_q)\}_{q=1}^Q$ are distinct, and the steering vectors on Array $\mathbb B$ $\{\bm{b}_0(\theta_q)\}_{q=1}^Q$ are distinct, too.

In (\ref{eq:VanA}), since the elements of $\bm \eta$ are non-zero, ${\rm diag}(\bm \eta)$ is a full rank diagonal matrix. In the Vandermonde matrix ${\bf V}_A$, each row vector  is the transposition of a steering vector of length ${L}_A-{K}_A+1$ and is distinct to one another. When ${L}_A-{K}_A+1 \geq Q$, the Vandermonde matrix has full row rank. Hence, ${\rm rank}({\bf W}_A)=Q$. Similarly, ${\rm rank}({\bf W}_B)=Q$ when ${L}_B-{K}_B+1 \geq Q$. $\bf W$ is henceforth a full rank matrix with rank $Q^2$. 

In (\ref{eq:PhiW}), the vectors $\{ \bm{a}_0(\theta_{q}) \otimes \bm{b}_0^*(\theta_{q}) \}_{q=1}^Q$ are $Q$ columns in the matrix $\bm{A}_0 \otimes \bm{B}_0^*$. When ${\bf W}{\bf W}^H$ has full rank, they are clearly in the signal subspace of $\bar{\bm \Phi}$.

{\it Case II}: The DOAs are ambiguous. For demonstrating purpose, we assume that two DOAs $\theta_1$ and $\theta_2$ satisfy $\theta_1 = \theta_2+2\pi m/M$ for a non-zero integer $m$. In this case, $\bm{a}_0(\theta_1) = \bm{a}_0(\theta_2)$ on Array $\mathbb A$. We also assume that the other steering vectors $\{\bm{a}_0(\theta_q)\}_{q=3}^Q$ are distinct.
In (\ref{eq:PhiKro}), we can delete the repeated column vector $\bm{a}_0(\theta_2)$ in $\bm{A}_0$ and the repeated row vector in ${\bf V}_A$, at the same time, combining the coefficients $\eta_1,\eta_2$ to obtain
\begin{IEEEeqnarray} {c} \label{eq:Delete}
\bm{A}_0 \cdot {\rm diag}(\bm{\eta}) \cdot {\bf V}_A = \left[ \bm{a}_0(\theta_1), \bm{a}_0(\theta_3), \ldots, \bm{a}_0(\theta_Q) \right] 
\nonumber \\ 
 \cdot {\rm diag}\left[ \eta_1+\eta_2, \eta_3,\ldots, \eta_Q \right]
\begin{bmatrix}
e^{jM0 \theta_1} & \ldots & e^{jM({L}_A-{K}_A) \theta_1} \\
e^{jM0 \theta_3} & \ldots & e^{jM({L}_A-{K}_A) \theta_3} \\
\vdots & \ddots & \vdots \\
e^{jM0 \theta_Q} & \ldots & e^{jM({L}_A-{K}_A) \theta_Q}
\end{bmatrix} \nonumber \\
= \tilde{\bm{A}}_0 \cdot {\rm diag}(\tilde{\bm{\eta}}) \cdot \tilde{\bf V}_A.
\end{IEEEeqnarray}
In (\ref{eq:Delete}), $\tilde{\bm A}_0$ is deleting the column vector $\bm{a}_0(\theta_2)$ from ${\bm A}_0$, the Vandermonde matrix $\tilde{\bf V}_A$ is deleting the row vector $\left[ e^{jM0 \theta_2}, \ldots, e^{jM({L}_A-{K}_A) \theta_2} \right]$ from the Vandermonde matrix ${\bf V}_A$. The remaining row vectors are distinct. The diagonal matrix ${\rm diag}(\tilde{\bm \eta})={\rm diag}\left[ \eta_1+\eta_2, \eta_3,\ldots, \eta_Q \right]$ is still full rank since $\eta_1+\eta_2 \neq 0$ from the non-vanishing assumption. Hence, the matrix ${\rm diag}(\tilde{\bm{\eta}}) \cdot \tilde{\bf V}_A$ has full rank.

Similarly, when the DOAs are ambiguous on Array $\mathbb B$, we can delete the repeated column vectors in $\bm{B}_0$ to obtain a full rank matrix $\tilde{\bm{B}}_0$, and delete the repeated row vectors in ${\bf V}_B$ to obtain a full rank matrix $ {\rm diag}(\tilde{\bm{\eta}}) \cdot \tilde{\bf V}_B$. Equation (\ref{eq:PhiKro}) becomes
\begin{IEEEeqnarray} {rCl}
\bm{\Upsilon} &=& \left( \tilde{\bm{A}}_0 \cdot {\rm diag}(\tilde{\bm{\eta}}) \cdot \tilde{\bf V}_A \right) \otimes 
\left( \tilde{\bm{B}}_0^* \cdot {\rm diag}(\tilde{\bm{\eta}}^*) \cdot \tilde{\bf V}_B^* \right) \nonumber \\
&=& \left( \tilde{\bm{A}}_0 \otimes \tilde{\bm{B}}_0^* \right) 
\left( {\rm diag}(\tilde{\bm{\eta}}) \cdot \tilde{\bf V}_A \right) \otimes  \left( {\rm diag}(\tilde{\bm{\eta}}^*) \cdot \tilde{\bf V}_B^* \right). \IEEEeqnarraynumspace
\end{IEEEeqnarray}

Analogous to Case I, the matrices ${\rm diag}(\tilde{\bm{\eta}}) \cdot \tilde{\bf V}_A$ and ${\rm diag}(\tilde{\bm{\eta}}) \cdot \tilde{\bf V}_B$ are of full rank. Since $\tilde{\bm A}_0$ and $\tilde{\bm B}_0$ are removing only the repeated steering vectors, for any DOA $\theta$, the vector $\bm{a}_0(\theta) \otimes \bm{b}_0^*(\theta)$ is still in the matrix $\tilde{\bm{A}}_0 \otimes \tilde{\bm{B}}_0^*$. 
Hence, the vectors $\{\bm{a}_0(\theta_q) \otimes \bm{b}_0(\theta_q)\}_{q=1}^Q$ are in the signal subspace of $\bar{\bm \Phi}$.

\ifCLASSOPTIONcaptionsoff
  \newpage
\fi
\bibliographystyle{IEEEtran}
\bibliography{library}

\end{document}